# Sub-Nanosecond Spin-Transfer Torque in an Ensemble of Superparamagnetic-Like Nanomagnets

Satoru Emori[1], Christoph Klewe[2], Jan-Michael Schmalhorst[3], Jan Krieft[3], Padraic Shafer[2], Youngmin Lim[1], David A. Smith[1], Arjun Sapkota[4,5], Abhishek Srivastava[4,5], Claudia Mewes[4,5], Zijian Jiang[1], Behrouz Khodadadi[1], Hesham Elmkharram[6], Jean J. Heremans[1], Elke Arenholz[2,7], Gunter Reiss[3], Tim Mewes[4,5]

1. Department of Physics, Virginia Tech, Blacksburg, VA 24061, USA

2. Advanced Light Source, Lawrence Berkeley National Laboratory, Berkeley, CA 94720, USA

3. Center for Spinelectronic Materials & Devices, Physics Department, Bielefeld University, Universitätsstraße 25, 33615 Bielefeld, Germany

4. Department of Physics and Astronomy, University of Alabama, Tuscaloosa, AL 35487, USA

5. Center for Materials for Information Technology (MINT), University of Alabama, Tuscaloosa, AL 35487, USA

6. Department of Materials Science and Engineering, Virginia Tech, Blacksburg, VA 24061, USA

7. Cornell High Energy Synchrotron Source, Ithaca, NY 14853, USA

**Spin currents can exert spin-transfer torques on magnetic systems even in the limit of vanishingly small net magnetization, as is the case for antiferromagnets. Here, we experimentally show that a spin-transfer torque is operative in a material with weak, short-range magnetic order – namely, a macroscopic ensemble of superparamagnetic-like Co nanomagnets. We employ element- and time-resolved X-ray ferromagnetic resonance (XFMR) spectroscopy to directly detect sub-ns dynamics of the Co nanomagnets, excited into precession with cone angle $\geq 0.003°$ by an oscillating spin current. XFMR measurements reveal that as the net moment of the ensemble decreases, the strength of the spin-transfer torque increases relative to those of magnetic field torques. Our findings point to spin-transfer torque as an effective way to manipulate the state of nanomagnet ensembles at sub-ns timescales.**

A flow of spin angular momentum, or spin current, injected into a thin-film magnetic medium can exert a spin-transfer torque (STT) on the magnetization[1–3]. STT enables a variety of scalable and energy-efficient nanoscale ferromagnetic devices for computing and communications applications[4–7]. Furthermore, STT can efficiently rotate the magnetic order of materials with zero net moment. For instance, STT (in particular, spin-orbit torque) allows for Néel vector switching[8,9] and auto-oscillations[10,11] in antiferromagnets. STT therefore may permit nanoscale information-technology devices based on antiferromagnets, which are insensitive to stray magnetic fields (e.g., of up to ~10 T) due to the strong inter-sublattice exchange coupling.

The net magnetization also averages to zero in a thermally disordered ensemble of weakly interacting ferromagnetic or superparamagnetic nanoparticles (e.g., often used in biomedical applications[12]), particularly in the absence of an applied magnetic field. These nanomagnets are not exchange-coupled to each other, such that a large fraction of the nanomagnet moments can relax (align) along a moderate field of ~0.1-1 T. However, this relaxation process involves a finite timescale, e.g., a few nanoseconds governed by the Gilbert damping rate[13]. On a shorter timescale, the moment $\mathbf{m_i}$ of each nanomagnet precesses about the field $\mathbf{H}$, as $\mathbf{m_i}$ is driven by the precessional torque $\boldsymbol{\tau_H} \sim -\mathbf{m_i} \times \mathbf{H}$. This field-driven precessional torque sums to zero in the limit of vanishing total magnetization (Fig. 1(a)), which is the case for a thermally disordered ensemble. By contrast, a spin current with polarization $\mathbf{s}$ exerts a STT of the form $\boldsymbol{\tau_{ST}} \sim \mathbf{m_i} \times \mathbf{s} \times \mathbf{m_i}$[1–3], which yields a finite sum even when the ensemble has zero net magnetization (Fig. 1(b)). Thus, on a sub-ns timescale, STT can yield a non-vanishing global torque in a nanomagnet ensemble with null net moment, whereas the precessional field torque alone cannot. This is partially analogous to the effectiveness of STT in antiferromagnets.

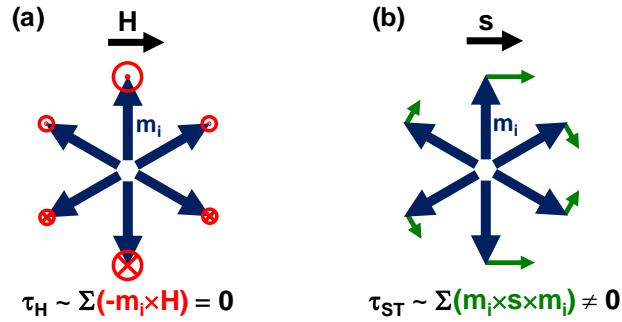

**Figure 1.** Illustrations of torques acting on an ensemble of magnetic moments, which sum to zero net magnetization, from (a) an externally applied field $\mathbf{H}$ and (b) spin current with polarization $\mathbf{s}$.

Prior experiments have shown that STT can control the state of a *single* superparamagnetic nanoisland[14] or nanoscale junction[7,15–17], as well as a nearly *saturated* ensemble of nanomagnets[18–20]. Yet, none has demonstrated STT in a macroscopic *ensemble* of nanomagnets in a *near-zero net magnetization* state (Fig. 1(b)). In this Letter, we present experimental confirmation of a global STT in such an ensemble of superparamagnetic-like nanomagnets. We perform spin pumping experiments[21–24] on a spin-valve-like film stack of NiFe/Cu/CoCu: the NiFe layer excited by microwave ferromagnetic resonance (FMR) pumps a coherent AC spin current that is absorbed by the granular CoCu spin sink, which consists of Co nanomagnets embedded in a nonmagnetic Cu matrix[25,26]. This nanomagnet ensemble is ferromagnetic-like at low temperature and superparamagnetic-like at room temperature, thereby allowing us to compare the effect of STT on these two distinct global magnetic states. We employ the element- and time-resolved X-ray ferromagnetic resonance (XFMR) technique[24,27–33] to directly detect torques on the Co nanomagnet ensemble at the sub-ns time scale. While torques from the microwave and interlayer dipolar fields decrease sharply in the superparamagnetic-like state, a substantial global STT generated by the AC spin current survives in the nanomagnet ensemble. Our results point to STT as an effective way to drive an ensemble of nanomagnets at the sub-ns time scale.

We employed DC sputter deposition with MgO substrates held at room temperature, resulting in polycrystalline films. Granular thin films of $Co_{25}Cu_{75}$ were grown by co-sputtering Co and Cu targets; Co and Cu are immiscible, such that nanoscale Co granules segregate in the matrix of Cu[25,26]. The film composition was set by the Co and Cu deposition rates and corroborated by energy-dispersive X-ray spectroscopy. We estimated an average granule size of <16 nm in $Co_{25}Cu_{75}$ films from powder X-ray diffractometry.

Single-layer 10-nm-thick $Co_{25}Cu_{75}$ films exhibit superparamagnetic-like behavior at room temperature. As shown in Fig. 2(a), our vibrating sample magnetometry measurements reveal room-temperature magnetization curves with zero coercivity and remanence. We observe similar magnetization curves for in-plane and out-of-plane field directions, indicating that static magnetic properties are not governed by the thin-film shape anisotropy. The nearly isotropic magnetization curves are consistent with isolated superparamagnetic-like Co granules embedded within the Cu matrix, rather than a homogeneous solid solution of Co and Cu atoms.

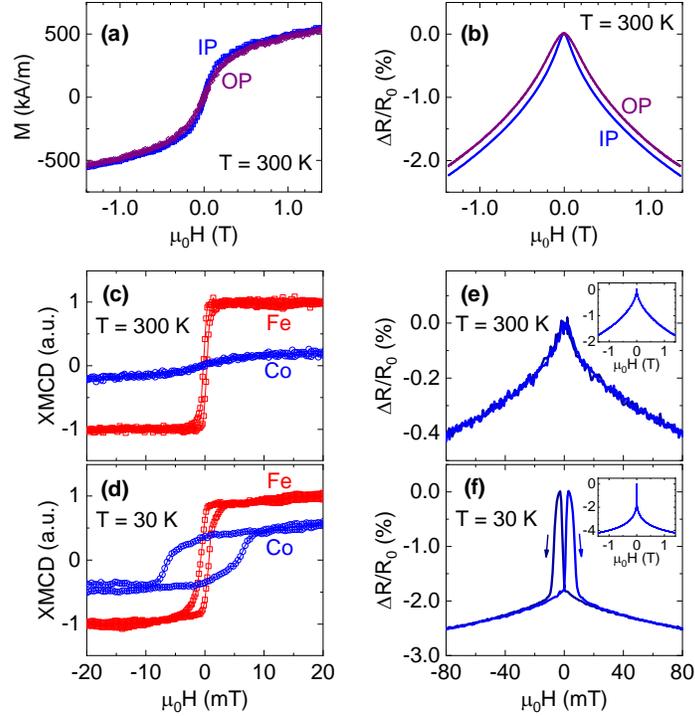

**Figure 2.** (a,b) Room-temperature in-plane (IP) and out-of-plane (OP) magnetization curves (a) and magnetoresistance curves (b) for single-layer $Co_{25}Cu_{75}(10)$. The magnetization in (a) is normalized by the estimated Co volume. (c,d) Element-resolved in-plane magnetization curves measured with XMCD for NiFe(10)/Cu(5)/CoCu(10) at (c) room temperature and (d) 30 K. (e,f) In-plane magnetoresistance curves for NiFe(10)/Cu(5)/CoCu(10) at (e) room temperature and (f) 30 K.

The magnetic field dependence of resistance (Fig. 2(b)) serves as additional evidence for the granular nature of the $Co_{25}Cu_{75}$ film. We observe a pronounced decrease in resistance $R$ with increasing magnitude of magnetic field, with a magnetoresistance ratio of $|R(0)-R(1.4\ \text{T})|/R(0) = |\Delta R|/R_0 \approx 2\%$ at room temperature. The magnetoresistance is similar for both in-plane and out-of-plane fields, consistent with previously reported isotropic giant magnetoresistance (GMR) in single-layer granular magnetic thin films[25,26].

We have further examined static magnetic properties of the granular $Co_{25}Cu_{75}$ film in a spin-valve-like $Ni_{80}Fe_{20}(10)/Cu(5)/Co_{25}Cu_{75}(10)$ stack (thickness unit: nm) designed for our spin pumping experiment. By utilizing element-resolved X-ray magnetic circular dichroism (XMCD), separate magnetization signals are obtained for the NiFe layer from the Fe $L_3$ edge and the CoCu layer from the Co $L_3$ edge. As shown in Fig. 2(c,d), the NiFe and CoCu layers show qualitatively distinct field dependence, which verifies that the two layers are not exchange coupled across the Cu spacer layer[34]. The room-temperature XMCD magnetization curve for CoCu shows superparamagnetic-like behavior with zero coercivity. By contrast, finite coercivity is observed at lower temperatures (e.g., 30 K, Fig. 2(d)), as thermal fluctuations are suppressed and the Co nanomagnets collectively behave like a ferromagnet. The room-temperature magnetoresistance curve of the NiFe/Cu/CoCu stack (inset Fig. 2(e)) is similar to that of single-layer CoCu (Fig. 2(b)) and indicates that the CoCu layer in the NiFe/Cu/CoCu stack is also granular. Low-temperature magnetoresistance curves show finite coercivity (Fig. 2(f)), consistent with the XMCD magnetization curve at the Co edge (Fig. 2(d)). Overall, our results in Fig. 2 corroborate the granular nature of $Co_{25}Cu_{75}$ and the superparamagnetic-like (ferromagnetic-like) behavior of the Co nanomagnet ensemble at room temperature (low temperature).

We now discuss the interplay of spin current and the Co nanomagnets in the NiFe/Cu/CoCu stack. We first look for evidence of the CoCu layer acting as a spin sink in broadband FMR spin pumping measurements[21–23], using a variable-temperature coplanar-waveguide spectrometer with the sample magnetized in the film plane. In these measurements, we detect and analyze the FMR signal from NiFe; the FMR signal from CoCu is negligibly small. From the linear slope of the NiFe FMR linewidth versus frequency (Fig. 3(a)), we obtain the Gilbert damping parameter $\alpha$ (see Supporting Information). At room temperature, $\alpha$ of the control sample without a CoCu layer is ≈0.007, in line with typical values for $Ni_{80}Fe_{20}$ (Refs. [35,36]).

Compared to this control sample, the NiFe/Cu/CoCu sample exhibits $\alpha$ that is enhanced by ≈0.002 (+30%). The magnitude of this damping enhancement is similar to prior results on spin-valve-like structures, where spin current is pumped from a NiFe layer and absorbed by another ferromagnetic layer[23]. The broadband FMR results thus suggest that granular CoCu acts as a sink for the spin current. We further observe that $\alpha$ is consistently greater by ≈0.002 for samples with the CoCu spin sink, independent of temperature (Fig. 3(b)).

However, the broadband FMR measurements do not indicate whether the spin current generates any STT in the Co nanomagnet ensemble. To probe the magnetization dynamics of the Co nanomagnets directly, we have performed time- and element-sensitive XFMR measurements under a continuous-wave 3-GHz microwave field excitation. Details of the XFMR method can be found in Supporting Information and Refs. 24,33, and here we emphasize that XFMR is a pump-probe technique that leverages XMCD to *separately* detect dynamics in the NiFe spin source (Fe $L_3$ edge) and the granular CoCu spin sink (Co $L_3$ edge). Specifically, we measured the oscillating magnetization (along the *y*-axis in Fig. 3(c)) transverse to the externally applied DC field $H_x$ (along the *x*-axis in Fig. 3(c)) for each Fe and Co.

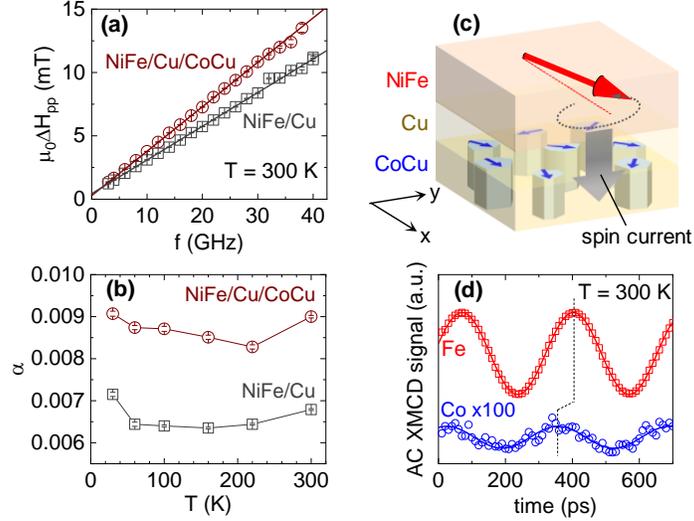

**Figure 3.** (a) Frequency dependence of the peak-to-peak FMR linewidth $\Delta H_{pp}$ for NiFe(10)/Cu(5)/CoCu(10) and control NiFe(10)/Cu(5) at room temperature. The solid lines show linear fits to obtain the Gilbert damping parameter. (b) Temperature dependence of the Gilbert damping parameter $\alpha$. (c) Schematic of FMR spin pumping, with NiFe as the spin source and Co nanomagnets as the spin sink. (d) Example of XFMR amplitude (AC XMCD) versus microwave delay for NiFe (Fe) and the nanomagnet spin sink (Co). The vertical dotted line emphasizes the offset in precessional phase.

Figure 3(d) shows examples of XFMR pump-probe delay scans, acquired at room temperature and $\mu_0 H_x = 9.6$ mT close to the resonance field of NiFe. Sinusoidal oscillations are evident for both the NiFe layer and the Co nanomagnets. We comment on two key observations: (1) Since the X-ray beam spot has a diameter of ~100 μm, the XFMR signal originates in the spatially averaged dynamics of $\gg 10^6$ Co nanomagnets. The observed sinusoidal oscillations for the Co nanomagnet ensemble, even when it is in the small-net-moment superparamagnetic-like state, shows strong evidence of the presence of a STT as we discuss below. (2) The Co magnetization precesses with a phase delay relative to the Fe magnetization, which implies that the dynamics of the Co nanogranules and the NiFe spin source are not directly coupled via static exchange interaction. Instead, the dynamics of Co and NiFe may be coupled via STT[21–24,29,33].

In addition to the STT, the microwave field[24] and the interlayer dipolar coupling field (e.g., Orange peel coupling)[27] could generate additional torques that drive the precession of the Co magnetization. Although these field torques vanish in systems with zero net magnetization (Fig. 1(a)), the net magnetization of the Co nanomagnet ensemble here is not strictly zero, due to the finite DC bias field of $\mu_0 H_x$ ~10 mT that is necessary for inducing the FMR of NiFe. We therefore must account for the possible roles of the microwave and dipolar field torques on the Co nanomagnets. On the other hand, we neglect a "field-like" STT, $\boldsymbol{\tau_{FLST}} \sim -\mathbf{m_i} \times \mathbf{s}$, which cannot be readily distinguished from the microwave and dipolar field torques. This assumption of negligible field-like STT is justified, because it is typically much smaller than the conventional "damping-like" or "Slonczewski-like" STT, $\boldsymbol{\tau_{ST}} \sim \mathbf{m_i} \times \mathbf{s} \times \mathbf{m_i}$, in metallic spin-valve-like stacks[1,2].

To determine the strength of the STT relative to the microwave and dipolar field torques, we analyze the amplitude and phase of magnetization precession versus $H_x$. Figure 4 summarizes our XFMR measurement results at 30 K (Fig. 4(a,b)) and 200 K (Fig. 4(c,d)) where the Co nanomagnet ensemble is ferromagnetic-like, and at room temperature (Fig. 4(e,f)) where the Co nanomagnets are superparamagnetic-like. The results show a clear FMR response of the NiFe spin source that is largely independent of temperature: the precessional amplitude, $A_{src} \propto \sqrt{\Delta H^2/[(H_x - H_{FMR})^2 + \Delta H^2]}$, exhibits a peak at the resonance field $\mu_0 H_{FMR} \approx 10$ mT with a half-width-at-half-maximum linewidth $\mu_0 \Delta H \approx 1$ mT, and the precessional phase,

$$\tan \phi_{src} = \Delta H/(H_x - H_{FMR}), \quad (1)$$

undergoes a shift of 180° across the resonance[28].

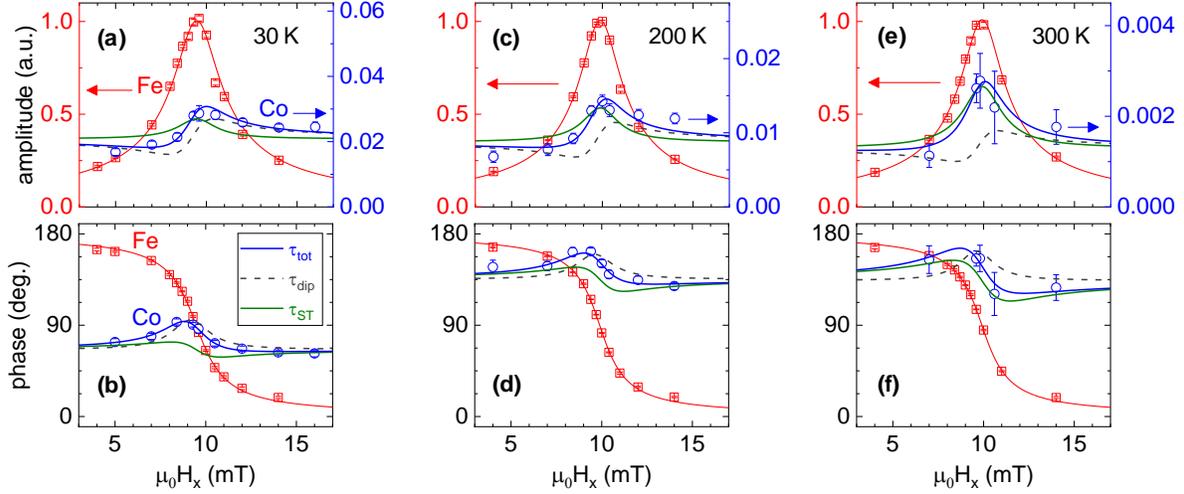

**Figure 4.** Field ($H_x$) dependence of precessional (a,c,e) amplitude and (b,d,e) phase for the NiFe spin source (Fe) and nanomagnet ensemble spin sink (Co) at (a,b) 30 K, (c,d) 200 K, and (e,f) room temperature. In each panel, the solid blue curve represents the fit with the total torque, $\tau_{tot}$, in the Co nanomagnet ensemble, taking into account both the interlayer dipolar torque ($\tau_{dip}$) and the STT ($\tau_{ST}$). The dashed gray curve represents the contribution from $\tau_{dip}$ (with $\beta_{ST} = 0$ in Eqs. (2) and (3)), and the solid green curve represents the contribution from $\tau_{ST}$ (with $\beta_{dip} = 0$ in Eqs. (2) and (3)).

The XFMR signal at the Co edge is more than an order of magnitude smaller, as shown in the plots of the Co amplitude normalized by the Fe amplitude (Fig. 4(a,c,e)). It was therefore impractical to acquire sufficient signal-to-noise ratios at many values of $H_x$ for Co within our allotted synchrotron beam time. Nevertheless, the data in Fig. 4 permit us to draw quantitative conclusions about the STT on the Co nanomagnets.

Firstly, the precessional phase for Co does not exhibit a 180° shift, which verifies the absence of Co FMR (i.e., the Co magnetization is not driven resonantly by the microwave field) in the measured range of $H_x$. We then self-consistently fit the observed amplitude $A^{Co}$ and phase $\phi^{Co}$ at the Co edge with the following equations, derived from coupled Landau-Lifshitz-Gilbert equations[24,31,33], accounting for the off-resonance microwave field torque, dipolar field torque, and STT:

$$A^{Co} = A_0^{Co}\sqrt{1 + (\beta_{dip}^2 + \beta_{ST}^2)\sin^2\phi_{src} + 2(\beta_{dip}\sin\phi_{src}\cos\phi_{src} + \beta_{ST}\sin^2\phi_{src})}, \quad (2)$$

$$\tan(\phi^{Co} - \phi_0^{Co}) = \frac{\beta_{dip}\sin^2\phi_{src} - \beta_{ST}\sin\phi_{src}\cos\phi_{src}}{1 + \beta_{dip}\sin\phi_{src}\cos\phi_{src} + \beta_{ST}\sin^2\phi_{src}}. \quad (3)$$

Here, $A_0^{Co}$ is a coefficient proportional to the microwave field torque, taken to be constant in the measured range of $H_x$. $\beta_{dip}$ and $\beta_{ST}$ are coefficients that parameterize the dipolar field torque and STT, respectively, normalized by the microwave field torque[24,33].

The dipolar field torque and STT are orthogonal to each other and hence exhibit qualitatively distinct $H_x$ dependences. For instance, the dipolar field torque yields a precessional amplitude that is antisymmetric about $H_x = H_{FMR}$ (dashed gray curve in Fig. 4(a,c,e)), whereas the STT yields a precessional amplitude that is symmetric about $H_x = H_{FMR}$ (solid green curve in Fig. 4(a,c,e)). This symmetry is reversed for the precessional phase (Fig. 4(b,d,f)): the dipolar torque (STT) generates a symmetric (antisymmetric) curve.

We further note that the microwave and dipolar field torques both depend on the net magnetization of the Co nanomagnet ensemble. As the net magnetization decreases with increasing temperature, the microwave and dipolar field torques are expected to decrease at the same rate. The coefficient $\beta_{dip}$ relating these two torques is thus assumed to be constant with temperature at $\beta_{dip}= 0.53+/-0.08$, derived from the results at 30 K (Fig. 4(a,b)). This simplifying assumption leaves $A_0^{Co}$ and $\beta_{ST}$ as the only free parameters in Eqs. (2) and (3) for fitting the 200 K and room temperature results.

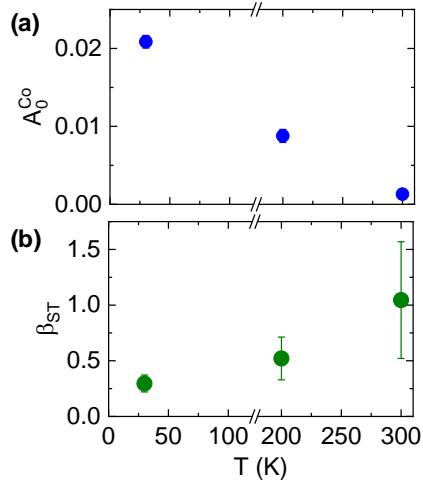

**Figure 5.** Temperature dependence of (a) $A_0^{Co}$, the coefficient proportional to the off-resonance microwave field torque, and (b) $\beta_{ST}$, coefficient proportional to the ratio between the STT and microwave field torque. The error bars are derived from the 95% confidence intervals of the fit parameters in Eqs. (2) and (3).

The amplitude of the Co XFMR signal decreases markedly with increasing temperature (Fig. 4(a,c,e)), as evidenced by an order of magnitude reduction in $A_0^{Co}$ from 30 K to room temperature (Fig. 5(a)). This trend is partially accounted for by the reduced net magnetization of the Co nanomagnet ensemble at higher temperatures. However, our XMCD magnetometry results (Fig. 2(c,d)) suggest that the Co net magnetization at $\mu_0 H_x \sim 10$ mT decreases by only a factor of ≈4 between 30 K and room temperature, such that there is likely an additional contribution to the ≈10-fold decrease of $A_0^{Co}$. We postulate that an increased effective damping in Co nanomagnets due to enhanced thermal fluctuations[37] may reduce the precessional cone angle for Co and hence $A_0^{Co}$.

While the net magnetization and the field torques in the nanomagnet ensemble become small at room temperature, an enhanced role of the STT relative to the field torques is suggested by the increase of $\beta_{ST}$ with increasing temperature, as shown in Fig. 5(b). Recalling that $\beta_{ST}$ is proportional to the ratio of the STT over the microwave field torque, the trend in Fig. 5(b) indicates that any reduction of the global STT in the nanomagnet ensemble is modest, compared to the sharp suppression of field torques, when magnetic order diminishes at elevated temperatures. This trend is also qualitatively consistent with the physical picture in Fig. 1 that the global STT remains finite even in a magnetic system with null net moment.

Furthermore, our results from different temperatures verify that STT is operative regardless of whether the Co nanomagnet spin sink is ferromagnetic-like or superparamagnetic-like: a coherent AC spin current generates a torque in each nanomagnet (e.g., superparamagnetic nanoparticle), resulting in a finite net torque summed over the macroscopic ensemble (Fig. 1(b)). Our findings thus point to STT as an effective mechanism at the sub-ns time scale to manipulate a macroscopic collection of superparamagnetic-like nanomagnets.

We finally comment on the sensitivity of the XFMR setup in our study. By comparing the amplitudes of the XFMR and static XMCD scans, we have estimated the resonant precessional cone angles. The cone angle for the FMR-driven NiFe spin source is ≈1.0º, similar to prior experiments[24,27–33]. Remarkably, the average cone angle of the Co nanomagnets at room temperature is estimated to be only ≈0.003º. This XFMR setup is therefore an excellent tool for examining small-angle dynamics in multi-layered and multi-element thin-film systems.

In summary, by employing time- and element-resolved XFMR spectroscopy, we have detected a STT that is driven by a coherent 3-GHz AC spin current in a macroscopic ensemble of Co nanomagnets. After disentangling the combined effects of the STT and the field torques, we find that the STT contribution is present regardless of whether the nanomagnet ensemble is in the low-temperature ferromagnetic-like state or the high-temperature superparamagnetic-like state. Our results highlight a fundamental feature of STT — that angular momentum supplied by a spin current can efficiently manipulate magnetic systems, even those with a vanishingly small net moment. From a practical perspective, STT may be an attractive mechanism to align an ensemble of nanomagnets for computing and sensing applications at sub-ns timescales.

This work used shared facilities at the Virginia Tech National Center for Earth and Environmental Nanotechnology Infrastructure (NanoEarth), a member of the National Nanotechnology Coordinated Infrastructure (NNCI), supported by NSF (ECCS 1542100). This research used resources of the Advanced Light Source, a U.S. DOE Office of Science User Facility under contract no. DE-AC02-05CH11231. A. Srivastava would like to acknowledge support through NASA grant CAN80NSSC18M0023 and A. Sapkota and C. M. would like to acknowledge support by NSF CAREER Award No. 1452670.

(1) Ralph, D. C.; Stiles, M. D. Spin Transfer Torques. *J. Magn. Magn. Mater.* **2008**, *320* (7), 1190–1216. https://doi.org/10.1016/j.jmmm.2007.12.019.

(2) Brataas, A.; Kent, A. D.; Ohno, H. Current-Induced Torques in Magnetic Materials. *Nat. Mater.* **2012**, *11* (5), 372–381. https://doi.org/10.1038/nmat3311.

(3) Manchon, A.; Železný, J.; Miron, I. M.; Jungwirth, T.; Sinova, J.; Thiaville, A.; Garello, K.; Gambardella, P. Current-Induced Spin-Orbit Torques in Ferromagnetic and Antiferromagnetic Systems. *Rev. Mod. Phys.* **2019**, *91* (3), 35004. https://doi.org/10.1103/RevModPhys.91.035004.

(4) Locatelli, N.; Cros, V.; Grollier, J. Spin-Torque Building Blocks. *Nat. Mater.* **2014**, *13* (1), 11–20. https://doi.org/10.1038/nmat3823.

(5) Sander, D.; Valenzuela, S. O.; Makarov, D.; Marrows, C. H.; Fullerton, E. E.; Fischer, P.;

McCord, J.; Vavassori, P.; Mangin, S.; Pirro, P.; et al. The 2017 Magnetism Roadmap. *J. Phys. D. Appl. Phys.* **2017**, *50* (36), 363001. https://doi.org/10.1088/1361-6463/aa81a1.

(6) Watanabe, K.; Jinnai, B.; Fukami, S.; Sato, H.; Ohno, H. Shape Anisotropy Revisited in Single-Digit Nanometer Magnetic Tunnel Junctions. *Nat. Commun.* **2018**, *9* (1), 663. https://doi.org/10.1038/s41467-018-03003-7.

(7) Borders, W. A.; Pervaiz, A. Z.; Fukami, S.; Camsari, K. Y.; Ohno, H.; Datta, S. Integer Factorization Using Stochastic Magnetic Tunnel Junctions. *Nature* **2019**, *573* (7774), 390–393. https://doi.org/10.1038/s41586-019-1557-9.

(8) Železný, J.; Gao, H.; Výborný, K.; Zemen, J.; Mašek, J.; Manchon, A.; Wunderlich, J.; Sinova, J.; Jungwirth, T. Relativistic Néel-Order Fields Induced by Electrical Current in Antiferromagnets. *Phys. Rev. Lett.* **2014**, *113* (15), 157201. https://doi.org/10.1103/PhysRevLett.113.157201.

(9) Wadley, P.; Howells, B.; Železný, J.; Andrews, C.; Hills, V.; Campion, R. P.; Novák, V.; Olejník, K.; Maccherozzi, F.; Dhesi, S. S.; et al. Electrical Switching of an Antiferromagnet. *Science.* **2016**, *351* (6273), 587.

(10) Cheng, R.; Xiao, D.; Brataas, A. Terahertz Antiferromagnetic Spin Hall Nano-Oscillator. *Phys. Rev. Lett.* **2016**, *116* (20), 207603. https://doi.org/10.1103/PhysRevLett.116.207603.

(11) Khymyn, R.; Lisenkov, I.; Tiberkevich, V.; Ivanov, B. A.; Slavin, A. Antiferromagnetic THz-Frequency Josephson-like Oscillator Driven by Spin Current. *Sci. Rep.* **2017**, *7*, 43705. https://doi.org/10.1038/srep43705.

(12) Neuberger, T.; Schöpf, B.; Hofmann, H.; Hofmann, M.; von Rechenberg, B. Superparamagnetic Nanoparticles for Biomedical Applications: Possibilities and Limitations of a New Drug Delivery System. *J. Magn. Magn. Mater.* **2005**, *293* (1), 483–

496. https://doi.org/10.1016/J.JMMM.2005.01.064.

(13) Mewes, C. K. A.; Mewes, T. Relaxation in Magnetic Materials for Spintronics. In *Handbook of Nanomagnetism: Applications and Tools*; Pan Stanford, 2015; pp 71–95.

(14) Krause, S.; Berbil-Bautista, L.; Herzog, G.; Bode, M.; Wiesendanger, R. Current-Induced Magnetization Switching with a Spin-Polarized Scanning Tunneling Microscope. *Science.* **2007**, *317* (5844), 1537–1540. https://doi.org/10.1126/science.1145336.

(15) Kiselev, S. I.; Sankey, J. C.; Krivorotov, I. N.; Emley, N. C.; Garcia, A. G. F.; Buhrman, R. A.; Ralph, D. C. Spin-Transfer Excitations of Permalloy Nanopillars for Large Applied Currents. *Phys. Rev. B* **2005**, *72* (6), 64430. https://doi.org/10.1103/PhysRevB.72.064430.

(16) Bapna, M.; Majetich, S. A. Current Control of Time-Averaged Magnetization in Superparamagnetic Tunnel Junctions. *Appl. Phys. Lett.* **2017**, *111* (24), 243107. https://doi.org/10.1063/1.5012091.

(17) Mizrahi, A.; Hirtzlin, T.; Fukushima, A.; Kubota, H.; Yuasa, S.; Grollier, J.; Querlioz, D. Neural-like Computing with Populations of Superparamagnetic Basis Functions. *Nat. Commun.* **2018**, *9* (1), 1533. https://doi.org/10.1038/s41467-018-03963-w.

(18) Chen, T. Y.; Huang, S. X.; Chien, C. L.; Stiles, M. D. Enhanced Magnetoresistance Induced by Spin Transfer Torque in Granular Films with a Magnetic Field. *Phys. Rev. Lett.* **2006**, *96* (20), 207203. https://doi.org/10.1103/PhysRevLett.96.207203.

(19) Luo, Y.; Esseling, M.; Münzenberg, M.; Samwer, K. A Novel Spin Transfer Torque Effect in $Ag_2$ Co Granular Films. *New J. Phys.* **2007**, *9* (9), 329–329. https://doi.org/10.1088/1367-2630/9/9/329.

(20) Wang, X. J.; Zou, H.; Ji, Y. Spin Transfer Torque Switching of Cobalt Nanoparticles. *Appl. Phys. Lett.* **2008**, *93* (16), 162501. https://doi.org/10.1063/1.3005426.

(21) Tserkovnyak, Y.; Brataas, A.; Bauer, G. Spin Pumping and Magnetization Dynamics in Metallic Multilayers. *Phys. Rev. B* **2002**, *66* (22), 224403. https://doi.org/10.1103/PhysRevB.66.224403.

(22) Heinrich, B.; Tserkovnyak, Y.; Woltersdorf, G.; Brataas, A.; Urban, R.; Bauer, G. E. W. Dynamic Exchange Coupling in Magnetic Bilayers. *Phys. Rev. Lett.* **2003**, *90* (18), 187601. https://doi.org/10.1103/PhysRevLett.90.187601.

(23) Ghosh, A.; Auffret, S.; Ebels, U.; Bailey, W. E. Penetration Depth of Transverse Spin Current in Ultrathin Ferromagnets. *Phys. Rev. Lett.* **2012**, *109* (12), 127202. https://doi.org/10.1103/PhysRevLett.109.127202.

(24) Li, J.; Shelford, L. R.; Shafer, P.; Tan, A.; Deng, J. X.; Keatley, P. S.; Hwang, C.; Arenholz, E.; van der Laan, G.; Hicken, R. J.; et al. Direct Detection of Pure Ac Spin Current by X-Ray Pump-Probe Measurements. *Phys. Rev. Lett.* **2016**, *117* (7), 76602. https://doi.org/10.1103/PhysRevLett.117.076602.

(25) Xiao, J. Q.; Jiang, J. S.; Chien, C. L. Giant Magnetoresistance in Nonmultilayer Magnetic Systems. *Phys. Rev. Lett.* **1992**, *68* (25), 3749–3752. https://doi.org/10.1103/PhysRevLett.68.3749.

(26) Berkowitz, A. E.; Mitchell, J. R.; Carey, M. J.; Young, A. P.; Zhang, S.; Spada, F. E.; Parker, F. T.; Hutten, A.; Thomas, G. Giant Magnetoresistance in Heterogeneous Cu-Co Alloys. *Phys. Rev. Lett.* **1992**, *68* (25), 3745–3748. https://doi.org/10.1103/PhysRevLett.68.3745.

(27) Arena, D. A.; Vescovo, E.; Kao, C.-C.; Guan, Y.; Bailey, W. E. Weakly Coupled Motion of Individual Layers in Ferromagnetic Resonance. *Phys. Rev. B* **2006**, *74* (6), 64409. https://doi.org/10.1103/PhysRevB.74.064409.

(28) Guan, Y.; Bailey, W. E.; Vescovo, E.; Kao, C.-C.; Arena, D. A. Phase and Amplitude of Element-Specific Moment Precession in Ni81Fe19. *J. Magn. Magn. Mater.* **2007**, *312* (2), 374–378. https://doi.org/10.1016/j.jmmm.2006.10.1111.

(29) Marcham, M. K.; Shelford, L. R.; Cavill, S. A.; Keatley, P. S.; Yu, W.; Shafer, P.; Neudert, A.; Childress, J. R.; Katine, J. A.; Arenholz, E.; et al. Phase-Resolved X-Ray Ferromagnetic Resonance Measurements of Spin Pumping in Spin Valve Structures. *Phys. Rev. B* **2013**, *87* (18), 180403. https://doi.org/10.1103/PhysRevB.87.180403.

(30) Stenning, G. B. G.; Shelford, L. R.; Cavill, S. A.; Hoffmann, F.; Haertinger, M.; Hesjedal, T.; Woltersdorf, G.; Bowden, G. J.; Gregory, S. A.; Back, C. H.; et al. Magnetization Dynamics in an Exchange-Coupled NiFe/CoFe Bilayer Studied by X-Ray Detected Ferromagnetic Resonance. *New J. Phys.* **2015**, *17* (1), 13019. https://doi.org/10.1088/1367-2630/17/1/013019.

(31) Baker, A. A.; Figueroa, A. I.; Love, C. J.; Cavill, S. A.; Hesjedal, T.; van der Laan, G. Anisotropic Absorption of Pure Spin Currents. *Phys. Rev. Lett.* **2016**, *116* (4), 47201. https://doi.org/10.1103/PhysRevLett.116.047201.

(32) Durrant, C. J.; Shelford, L. R.; Valkass, R. A. J.; Hicken, R. J.; Figueroa, A. I.; Baker, A. A.; van der Laan, G.; Duffy, L. B.; Shafer, P.; Klewe, C.; et al. Dependence of Spin Pumping and Spin Transfer Torque upon $Ni_{81}Fe_{19}$ Thickness in Ta / Ag / $Ni_{81}Fe_{19}$ / Ag / $Co_2MnGe$ / Ag / Ta Spin-Valve Structures. *Phys. Rev. B* **2017**, *96* (14), 144421. https://doi.org/10.1103/PhysRevB.96.144421.

(33) Li, Q.; Yang, M.; Klewe, C.; Shafer, P.; N'Diaye, A. T.; Hou, D.; Wang, T. Y.; Gao, N.; Saitoh, E.; Hwang, C.; et al. Coherent Ac Spin Current Transmission across an Antiferromagnetic CoO Insulator. *Nat. Commun.* **2019**, *10* (1), 5265.

https://doi.org/10.1038/s41467-019-13280-5.

(34) Leal, J. L.; Kryder, M. H. Oscillatory Interlayer Exchange Coupling in Ni$_{81}$Fe$_{19}$/Cu/Ni$_{81}$Fe$_{19}$/Fe$_{50}$Mn$_{50}$ Spin Valves. *J. Appl. Phys.* **1996**, *79* (5), 2801–2803. https://doi.org/10.1063/1.361115.

(35) Schoen, M. A. W.; Lucassen, J.; Nembach, H. T.; Silva, T. J.; Koopmans, B.; Back, C. H.; Shaw, J. M. Magnetic Properties in Ultrathin 3d Transition-Metal Binary Alloys. II. Experimental Verification of Quantitative Theories of Damping and Spin Pumping. *Phys. Rev. B* **2017**, *95* (13), 134411. https://doi.org/10.1103/PhysRevB.95.134411.

(36) Zhao, Y.; Song, Q.; Yang, S.-H.; Su, T.; Yuan, W.; Parkin, S. S. P.; Shi, J.; Han, W. Experimental Investigation of Temperature-Dependent Gilbert Damping in Permalloy Thin Films. *Sci. Rep.* **2016**, *6*, 22890. https://doi.org/10.1038/srep22890.

(37) Chubykalo-Fesenko, O.; Nowak, U.; Chantrell, R. W.; Garanin, D. Dynamic Approach for Micromagnetics close to the Curie Temperature. *Phys. Rev. B* **2006**, *74* (9), 94436. https://doi.org/10.1103/PhysRevB.74.094436.